# Generative chemistry: drug discovery with deep learning generative models

Yuemin Bian[1,2] and Xiang-Qun Xie[1,2,3,4]*

[1]Department of Pharmaceutical Sciences and Computational Chemical Genomics Screening Center, School of Pharmacy; [2]NIH National Center of Excellence for Computational Drug Abuse Research; [3]Drug Discovery Institute; [4]Departments of Computational Biology and Structural Biology, School of Medicine, University of Pittsburgh, Pittsburgh, Pennsylvania 15261, United States.

*Corresponding Author: Xiang-Qun Xie, MBA, Ph.D.
Professor of Pharmaceutical Sciences/Drug Discovery Institute
Director of CCGS and NIDA CDAR Centers
335 Sutherland Drive, 206 Salk Pavilion
University of Pittsburgh
Pittsburgh, PA15261, USA
412-383-5276 (Phone)
412-383-7436 (Fax)
Email: xix15@pitt.edu

**Abstract**:

The *de novo* design of molecular structures using deep learning generative models introduces an encouraging solution to drug discovery in the face of the continuously increased cost of new drug development. From the generation of original texts, images, and videos, to the scratching of novel molecular structures, the incredible creativity of deep learning generative models surprised us about the height machine intelligence can achieve. The purpose of this paper is to review the latest advances in generative chemistry which relies on generative modeling to expedite the drug discovery process. This review starts with a brief history of artificial intelligence in drug discovery to outline this emerging paradigm. Commonly used chemical databases, molecular representations, and tools in cheminformatics and machine learning are covered as the infrastructure for the generative chemistry. The detailed discussions on utilizing cutting-edge generative architectures, including recurrent neural network, variational autoencoder, adversarial autoencoder, and generative adversarial network for compound generation are focused. Challenges and future perspectives follow.

# 1. INTRODUCTION

Drug discovery is expensive. The average cost for the development of a new drug now hits 2.6 billion USD and the overall discovery process takes over 12 years to finish[1, 2]. Moreover, these numbers keep increasing. It is critical to think and explore efficient and effective strategies to confront the growing cost and to accelerate the discovery process. The progression in the high-throughput screening (HTS) dramatically speeded up the task of lead identification by screening candidate compounds in large volume[3, 4]. When it comes to the lead identification, the concept can be further classified into two divisions, the structure-based approach[5, 6] and the ligand-based approach[7]. Combining with the significant progress in computation, the development of these two approaches has resulted in constructive virtual screening (VS) methodologies. Traditionally, with the structure of the target protein available, structure-based approaches including molecular docking studies[8-10], molecular dynamic simulations[11-13], fragment-based approach[10, 14], etc. can be applied to explore the potential receptor-ligand interactions and to virtually screen a large compound set for finding the plausible lead. Then with the identified active molecules for the given target, ligand-based approaches such as pharmacophore modeling[15, 16], scaffolding hopping[17, 18], and molecular fingerprint similarity search[19] can be conducted for modifying known leads and for finding future compounds. The rapid advancement in computational power and the blossom of machine learning (ML) algorithms brought the ML-based decision-making model[20, 21] as an alternative path to the VS campaigns in the past decades. There is increased availability of data in cheminformatics and drug discovery. The capability of dealing with large data to detect hidden patterns and to facilitate the future data prediction in a time-efficient manner favored ML in building VS pipelines.

It is encouraging to note the successful applications of the above-mentioned computational chemistry approaches and ML-based VS pipelines on drug discovery these days. The conventional methods are effective. However, the challenge remains on developing pioneering methods, techniques, and strategies in the confrontation of the costly procedure of drug discovery. The flourishing of deep learning generative models brings fresh solutions and opportunities to this field. From the generated human faces that are indistinguishable with real people[22], to the text generation tools that mimic the tone and vocabulary of certain authors[23], the astonishing creativity of deep learning generative models brings our understanding of the machine intelligence to a new level. In recent years, the expeditions toward generative chemistry mushroomed, which explored the possibility of utilizing generative models to effectively and efficiently design molecular structures with desired properties. Promising and compelling outcomes including the identification of DDR1 kinase inhibitors within 21 days using deep learning generative models[24] may indicate that we are probably at the corner of an upcoming revolution of drug discovery in the artificial intelligence (AI) era. This review article starts with a brief evolution of AI in drug discovery, and the infrastructures in both cheminformatics and machine learning. The state-of-the-art generative models including recurrent neural networks (RNNs), variational autoencoders (VAEs), adversarial autoencoders (AAEs), and generative adversarial networks (GANs) are focused on to discuss their fundamental architectures as well as their applications in the *de novo* drug design.

## 2. ARTIFICIAL INTELLIGENCE IN DRUG DISCOVERY

Artificial intelligence (AI) is the study of developing and implementing techniques that enable the machine to behave with human-like intelligence[25]. The concept of AI can be traced back to the 1950s when researches questioned whether computers can be made to handle automated intelligence tasks which are commonly fulfilled by humans[26]. Thus, AI is a broad area of research that includes both (1) methodologies employing learning processes and (2) approaches that no learning process is involved in. At the early stage, researchers believed that by defining a sufficient number of explicit rules to maneuver knowledge, the human-level AI can be expected (**Fig. 1a**). In the face of a specific problem, the human studying process on existing observations can contribute to the accumulation of knowledge. Explicit rules were expected to describe knowledge. By programming and applying these rules, the answers for future observations are anticipated. This strategy is also known as symbolic AI[27]. Symbolic AI is an efficient solution to logical problems, for instance, chess playing. However, when handling problems with blurry, unclear, and distorted knowledge, such as image recognition, language translation, and to our topic, the classification of active compounds from decoys for a therapeutic target, symbolic AI turned out to show limited capability. We may define explicit rules to guide the selection of general drug-like compounds, Lipinski's rule of five[28] for example, but it is almost impossible to exhaust specified rules for guiding the selection of agonists to cannabinoid receptor 2 or other targets[29]. Machine learning (ML) took over symbolic AI's position as a novel method with the ability to learn on its own.

ML allows computers to solve specific tasks by learning on their own[30, 31]. Through directly looking at the data, computers can summarize the rules instead of waiting for programmers to craft them (**Fig. 1b**). In the paradigm of ML-based problem solving, data and the answers to the data are functioned as input with rules as the outcome. The produced rules can then be applied to predict answers for future data. Statistical analysis is always associated with ML, while they can be distinguished at several aspects[32]. The application of ML is usually towards large and complex datasets, such as a dataset with millions of small molecules that cover a huge chemical space with diversified scaffolds, which statistical analysis can be incapable to deal with. The flourish of ML starts in the 1990s[33]. The method rapidly became a dominant player in the field of AI. Commonly used ML systems in drug discovery can be categorized into supervised learning, unsupervised learning, and reinforcement learning (**Fig. 1c**). In supervised learning, the algorithms are fed with both the data and the answers to these data (label). Protein family subtypes selectivity prediction is an example for classification: the classifier is trained with numbers of sample molecules along with their labels (the specific protein family member they interact with), and the well-trained classifier should be able to classify the future molecules[20, 29, 34, 35]. Quantitative structure-activity relationship analysis is an example for regression: the regressor is trained with molecules sharing similar scaffold along with their biological activity data (*Ki*, *IC$_{50}$*, and *EC$_{50}$* values for example), and the well-trained regressor should be able to predict the numeric activity values for future molecules with the similar scaffold[10, 36]. In unsupervised learning, the algorithms are trained with unlabeled data. For instance, a high-throughput screening campaign may preselect a smaller representative compound set from a large compound database using the clustering method to group molecules with similar structures into clusters[37, 38]. A subset of molecules selected from different clusters can then offer improved structural diversity to cover a bigger chemical space than random pickup. In reinforcement learning, the learning system can choose actions according to its observation of the environment, and get a penalty (or reward) in return[39]. To achieve the lowest penalty (or highest reward), the system must learn and choose the best strategy by itself.

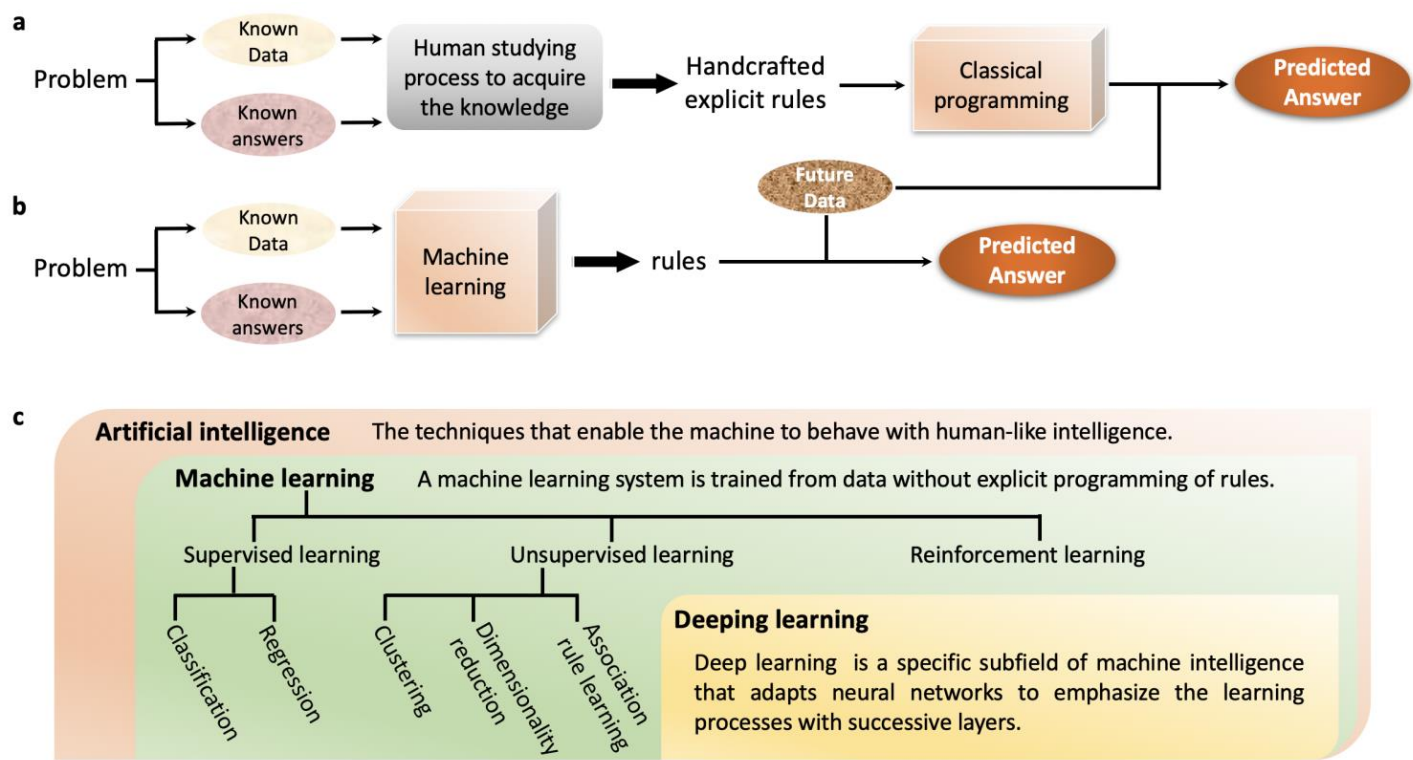

**Figure 1. From artificial intelligence to deep learning. a. The programming paradigm for symbolic AI. b. The programming paradigm for ML. c. The relationship among artificial intelligence, machine learning, and deep learning.**

Deep learning (DL) is a specific subfield of ML that adapts neural networks to emphasize the learning processes with successive layers (**Fig. 1c**). DL methods can transfer the representation at one level to a higher and more abstract level[40]. The feature of representation learning enables DL methods to discover representations from the raw input data for tasks such as detection and classification. The word "deep" in DL reflects this character of successive layers of representations, and the number of layers determines the depth of a DL model[41]. In contrast, conventional ML methods that transform the input data into one or two successive representation spaces are sometimes referred to as shallow learning methods. The vast development in the past decades brought DL great flexibility on the selection of architectures, such as the fully connected artificial neural network (ANN) or multi-layer perceptron (MLP)[42], convolutional neural network (CNN)[43], and recurrent neural network (RNN)[44]. The rise of generative chemistry is largely benefited from the extensive advancement of generative modeling, which predominantly depends on the flourishing of DL architectures. The successful application of the Long Short-Term Memory (LSTM) model[45], a special type of RNN model, on text generation inspired the simplified molecular-input line-entry system (SMILES)-based compound design. And the promising exercise of using the Generative Adversarial Network (GAN) model[46] for image generation motivated the fingerprint and graph centered molecular structural scratch. The major reason for DL to bloom rapidly can be that the very method provides solutions to previously unsolvable problems and outperforms the competitors with simplified representation learning process[26, 40]. It is foreseen that the process of molecule design can evolve into a more efficient and effective manner with the proper fusion with DL.

## 3. DATA SOURCES AND MACHINE LEARNING INFRASTRUCTURES

Deep learning campaigns start with high-quality input data. The successful development of generative chemistry models relies on cheminformatics and bioinformatics data for the molecules and biological systems. **Table 1** exhibits some routinely used databases in drug discovery for both small and large biological molecules. In a typical case of structure-based drug discovery, a 3D model of the protein (or DNA/RNA) target is critical for the following steps on evaluating potential receptor-ligand interactions. PDB database[47] is a good source for accessing structural information for large biological systems, and the UniProt database[48] will be a convenient source for sequence data. Regarding chemicals, PubChem[49] can be a go-to place. PubChem is comprehensive. It currently contains ~103 million compounds (with unique chemical structures) and ~253 million substances (information about chemical entities). If the major focus is on bioactive molecules, ChEMBL[50] can be an efficient database to interact. ChEMBL currently documents ~2 million reported drug-like compounds with bioactivity data for 13,000 targets. Supposing that the interest is more on studying the existing drugs on the market instead of drug-like compounds, the DrugBank[51] can serve. To date, DrugBank records ~ 14,000 drugs, including approved small molecule drugs and biologics, nutraceuticals, and discovery-phase drugs. With virtual screening campaigns, adding some commercially available compounds to in-house libraries are preferred as they may further increase the structural diversity and expand the coverage of the chemical space. Once potential hits were predicted to be among these compounds, the commercial availability gives them easy access for future experimental validations. Zinc database[52] now archives ~230 million purchasable compounds in ready-to-dock format. It is worth mentioning that constructing topic-specific and target-specific databases is trending. ASD[53] is one example that files allosteric modulators and related macromolecules to facilitate the research on allosteric modulation. The rising of Chemogenomics databases[54, 55] for certain diseases and druggable targets is another example that these libraries focus on particular areas of research.

With the input data ready, the next consideration is transforming the data into machine-readable format. **Table 2** lists commonly used molecular representations. SMILES[56] describes molecular structures in a text-based format using short ASCII strings. Multiple SMILES strings can be generated for the same molecule with different starting atoms. This ambiguity led to the effects of canonicalization that determines which of all possible SMILES will be used as the reference SMILES for a molecule. Popular cheminformatics packages such as OpenEye[57] and RDKit[58] are possible solutions for standardizing the canonical SMILES[59]. The canonical SMILES is a type of well-liked molecular representation in generative chemistry models as it packs well with language processing and sequence generation techniques like RNNs. Usually the SMILES strings are first converted with one-hot encoding. The categorical distribution for each element can then be produced by the generative models. Fingerprints are another vital group of molecular representations. Molecular Access System (MACCS) fingerprint[60] has 166 binary keys, each of which indicates the presence of one of the 166 MDL MACCS structural keys calculated from the molecular graph. Fingerprints can be calculated through different approaches. By enumerating circular fragments, linear fragments, and tree fragments from the molecular graph, Circular[61], Path, and Tree fingerprints[62] can be created. Using fingerprints as representations may suffer from inconvertibility in that the complete structure of a molecule cannot be reconstructed directly from the fingerprints[63]. To have fingerprints calculated for a large enough compounds library to function as a look-up index may be a compromised solution[64]. Despite this difficulty, fingerprints are popular among ML classification models for tasks like distinguishing active compounds from inactive ones for a given target.

**Table 1. Well established cheminformatics databases available for drug discovery**

| Database | Description | Web linkage | Examples of usage |
|---|---|---|---|
| **UniProt[48]** | The Universal Protein Resource (UniProt) is a resource for protein sequence and annotation data | https://www.uniprot.org | Protein sequence homology search, alignment, and protein ID retrieving especially for structural-based drug discovery |
| **RCSB PDB[47]** | The Protein Data Bank (PDB) provides access to 3D structure data for large biological molecules, including protein, DNA, and RNA. | https://www.rcsb.org | Protein 3D structures are fundamental for hot spot identification, docking simulation, and molecular dynamics simulation in structural-based drug discovery. |
| **PDBbind[65]** | PDBbind provides a collection of the experimentally measured binding affinity data for all types of biomolecular complexes deposited in the PDB. | http://www.pdbbind.org.cn | The receptor-ligand binding data for resolved protein structures can function as the benchmark to evaluate future simulations |
| **PubChem[49]** | PubChem is the world's largest collection of chemical information. | https://pubchem.ncbi.nlm.nih.gov | To acquire comprehensive chemical information ranging from NMR spectra, physical-chemical properties, to biomolecular interactions. |
| **ChEMBL[50]** | ChEMBL is a manually curated database of bioactive molecules with drug-like properties. | https://www.ebi.ac.uk/chembl/ | To collect cheminformatics data of reported molecules for a given target. A high-quality compound collection is the key to the ligand-based drug discovery |
| **SureChEMBL[66]** | SureChEMBL is a resource containing compounds extracted from patent literature. | https://www.surechembl.org/search/ | Compound-patent associations |
| **BindingDB[67]** | BindingDB is a database of measured binding affinities for the interactions of protein considered to be drug-targets with small, drug-like molecules. | https://www.bindingdb.org/bind/index.jsp | To retrieve compound sets for a specific target similar to ChEMBL but with the focus on experimental binding affinities. |
| **DrugBank[51]** | The DrugBank database combines detailed drug data with comprehensive drug target information | https://www.drugbank.ca | Drug repurposing study for existing drugs. On-target and off-target analysis for a compound. |
| **ZINC[52]** | Zinc is a database of commercially-available compounds | https://zinc.docking.org | Zinc database is good for virtual screening on hit identification as the compounds are commercially available for quick biological validations afterwards. |
| **Enamine** | Enamine provides an enumerated database of synthetically feasible molecules | https://enamine.net | The establishment of a target-specific compound library. Fragment-based drug discovery. |
| **ASD[53]** | Allosteric Database (ASD) provides a resource for structure, function, disease and related annotation for allosteric macromolecules and allosteric modulators | http://mdl.shsmu.edu.cn/ASD/ | To facilitate the research on allosteric modulation with enriched chemical data on allosteric modulators. |
| **GDB[68]** | GDB databases provide multiple subsets of combinatorially generated compounds following chemical stability and synthetic feasibility rules | http://gdb.unibe.ch/downloads/ | Using combinatorial chemistry is a good way to largely expand the chemical space. |

**Table 2. Examples of commonly used molecular representations**

| | Representation | Description |
|---|---|---|
| | **SMILES**[56] | The simplified molecular-input line-entry system (SMILES) is a specification in the form of a line notation for describing the structure of chemical species using short ASCII strings. |
| | **Canonical SMILES** | Canonicalization is a way to determine which of all possible SMILES will be used as the reference SMILES for a molecular graph. |
| | **InChI**[69] | The International Chemical Identifier (InChI) is a textual identifier for chemical substances, designed to provide a standard way to encode molecular information. |
| | **InChI Key** | The condensed, 27 character InChI Key is a hashed version of the full InChI. |
| **Fingerprints** | **MACCS Keys**[60] | MACCS keys are 166 bit structural key descriptors in which each bit is associated with a SMARTS pattern. |
| | **Circular**[61, 70] | Circular fingerprints are created by exhaustively enumerating all circular fragments grown radially from each heavy atom of the molecule up to the given radius. |
| | **Path**[62] | Path fingerprints are created by exhaustively enumerating all linear fragments of a molecular graph up to a given size. |
| | **Tree**[62] | Tree fingerprints are generated by exhaustively enumerating all tree fragments of a molecular graph up to a given size. |
| | **Atom Pair**[71] | Atom Pair fingerprints encode each atom as a type, enumerates all distances between pairs, and then hashes the results. |

After collecting the high-quality data and transforming the data into the appropriate format, it is time to apply data science to the development of the predictive models. **Table 3** illustrates examples of frequently considered cheminformatics toolkits and machine learning packages. RDKit, Open Babel[72], and CDK[73] are cheminformatics toolkits that are comprised of a set of libraries with source codes for various functions, such as chemical files I/O formatting, substructure and pattern search, and molecular representations generation. The typical applications of deploying these toolkits can contribute to virtual screening, structural similarity search, structure-activity relationship analysis, etc[74]. The workflow environment is not unique to the cheminformatics research, but can facilitate the automation of data processing with a user-friendly interface. The workflow systems like KNIME[75, 76] can execute tasks in succession and perform recurring tasks efficiently, such as iterative fingerprints calculation for a compound library. The strategy of integrating cheminformatics toolkits as nodes into a workflow and connecting them with edges is gaining popularity and is increasingly employed[77-79]. When it comes to ML and DL modeling, TensorFlow[80], CNTK[81], Theano[82], and PyTorch[83] are well-recognized packages for employment. These packages handle low-level operations including tensor manipulation and differentiation. In contrast, Keras[84] is a model-level library that deals with tasks in a modular way. As a high-level API, Keras is running on top of TensorFlow, CNTK, and Theano. Scikit-Learn[85] is an efficient and straightforward tool for predictive data analysis. It is known more for its role in conventional ML modeling as the library comprehensively integrates algorithms like Support Vector Machine (SVM), Random Forest (RF), Logistic regression, Naïve Bayes (NB), etc.

**Table 3. Commonly used cheminformatics and machine learning packages**

| Package | Description | Web linkage |
| --- | --- | --- |
| RDKit[58] | RDKit is an open-source toolkit for cheminformatics. Features include 2D and 3D molecular operations, descriptor generation, molecular database cartridge, etc. | https://www.rdkit.org |
| Open Babel[72] | Open Babel is an open chemical toolbox to search, convert, analyze, or store data from molecular modeling, chemistry, solid-state materials, biochemistry, or related areas. | http://openbabel.org/wiki/Main_Page |
| CDK[73] | The Chemistry Development Kit (CDK) is a collection of modular Java libraries for processing cheminformatics. | https://cdk.github.io |
| KNIME[75] | KNIME is a workflow environment in data science that can be integrated to automate certain cheminformatics operations. | https://www.knime.com |
| TensorFlow[80] | TensorFlow is an open-source platform for machine learning. It has a set of tools, libraries, and community resources that enable researchers to build and deploy ML applications. | https://www.tensorflow.org |
| CNTK[81] | The Cognitive Toolkit (CNTK) is an open-source toolkit for commercial-grade distributed deep learning. It describes neural networks as a series of computational steps via a directed graph. | https://github.com/microsoft/CNTK |
| Theano[82] | Theano is a Python library for defining, optimizing, and evaluating mathematical expressions. | http://deeplearning.net/software/theano/ |
| PyTorch[83] | PyTorch is an open-source machine learning library based on the Torch library. | https://pytorch.org |
| Keras[84] | Keras is a high-level neural networks API, written in Python and capable of running on top of TensorFlow, CNTK, or Theano. It was developed with a focus on enabling fast experimentation. | https://keras.io |
| Scikit-Learn[85] | Scikit-learn is a free software machine learning library for the Python programming language. | https://scikit-learn.org/stable/ |

## 4. GENERATIVE CHEMISTRY WITH THE RECURRENT NEURAL NETWORK (RNN)

RNN is a widely used neural network architecture in generative chemistry for proposing novel structures. As a type of powerful generative model especially in natural language processing, RNNs usually use sequences of words, strings, or letters as the input and output[44, 86-88]. In this case, the SMILES strings are usually employed as a molecular representation. Different from ANNs and CNNs which do not have memories, RNNs iteratively process sequences and store a state holding current information. On the contrary, ANNs and CNNs process each input independently without stored information between them. When describing an RNN, it can be considered as a network with an internal loop that loops over the sequence elements instead of processing in a single step (**Fig. 2a**). The state that stored information will be updated during each loop. For simplicity, the process of

computing the output $y$ can follow the equation: $y$ = activation (W$_o$$x$ + U$_o$$h$ + b$_o$), where W$_o$ and U$_o$ are weight matrices for the input $x$ and state $h$, and b$_o$ as a bias vector. **Figure 2a** can represent the structure of a simple RNN model. However, this structure can suffer severely from the vanishing gradient problem which makes neural networks untrainable after adding more layers. Even though the state $h$ is supposed to hold the information from the sequence elements previously seen, the long-term dependencies make the learning process impossible[89, 90]. The Long Short-Term Memory (LSTM) algorithm[45] was developed to overcome this shortcoming. The LSTM layer attaches a carry track to carry information across the learning process to counter the loss of signals from gradual vanishing (**Fig. 2b**). With this carry track, the information learned from each sequence element can be loaded, and the loaded information can be transported and accessed at a later stage. The process of computing the output $y$ for LSTM is similar with the previous equation but adding the contribution of the carry track: $y$ = activation (W$_o$$x$ + U$_o$$h$ + V$_o$$c$ + b$_o$), where W$_o$, U$_o$, and V$_o$ are weight matrices for the input $x$, state $h$, and carry $c$, and b$_o$ as a bias vector. In certain cases, multiple recurrent layers in a model can be stacked to enhance representational power.

A typical framework on generative modeling for molecule generation applying LSTM algorithm (**Fig. 2c**) starts with the collection of training molecules. The RNN model can be fine-tuned through the transfer learning that first accumulates knowledge from the large compound datasets and then produces the novel structures by learning smaller focused datasets. When the collections of training molecules (for large sets or small focused sets) are ready, SMILES strings can be calculated for each molecule. One-hot encoding is a regular operation for processing the molecular representations. In one-hot encoding, a unique integer index $i$ is assigned to every character in the SMILES string. Then a binary vector can be constructed of size $C$ (the number of unique characters in the string) with all zeros but for the $i$th entry which is one. For instance, there are four ($C = 4$) unique characters, "C", "N", "c", and "1" in SMILES strings, input "C" is transferred to (1, 0, 0, 0), "N" to (0, 1, 0, 0), "c" to (0, 0, 1, 0), and "1" to (0, 0, 0, 1) after one-hot encoding. In practice, usually an additional starting character like "G" and an ending character like "E" will be added to the SMILES to denote a complete string. The neural network with LSTM layer(s) can be trained to predict the $n+1$th character given the input of string with $n$ characters. The probability of distribution for the $n+1$th character is calculated as the loss to evaluate the model performance. With the trained model, the sampling process can start with the starting character or certain SMILES strings of molecular fragments to sample the next character until the ending character is hit. The SMILES strings are reversed from the generated binary matrices according to the previous one-hot encoding to construct the molecular graphs as the output for this generative model.

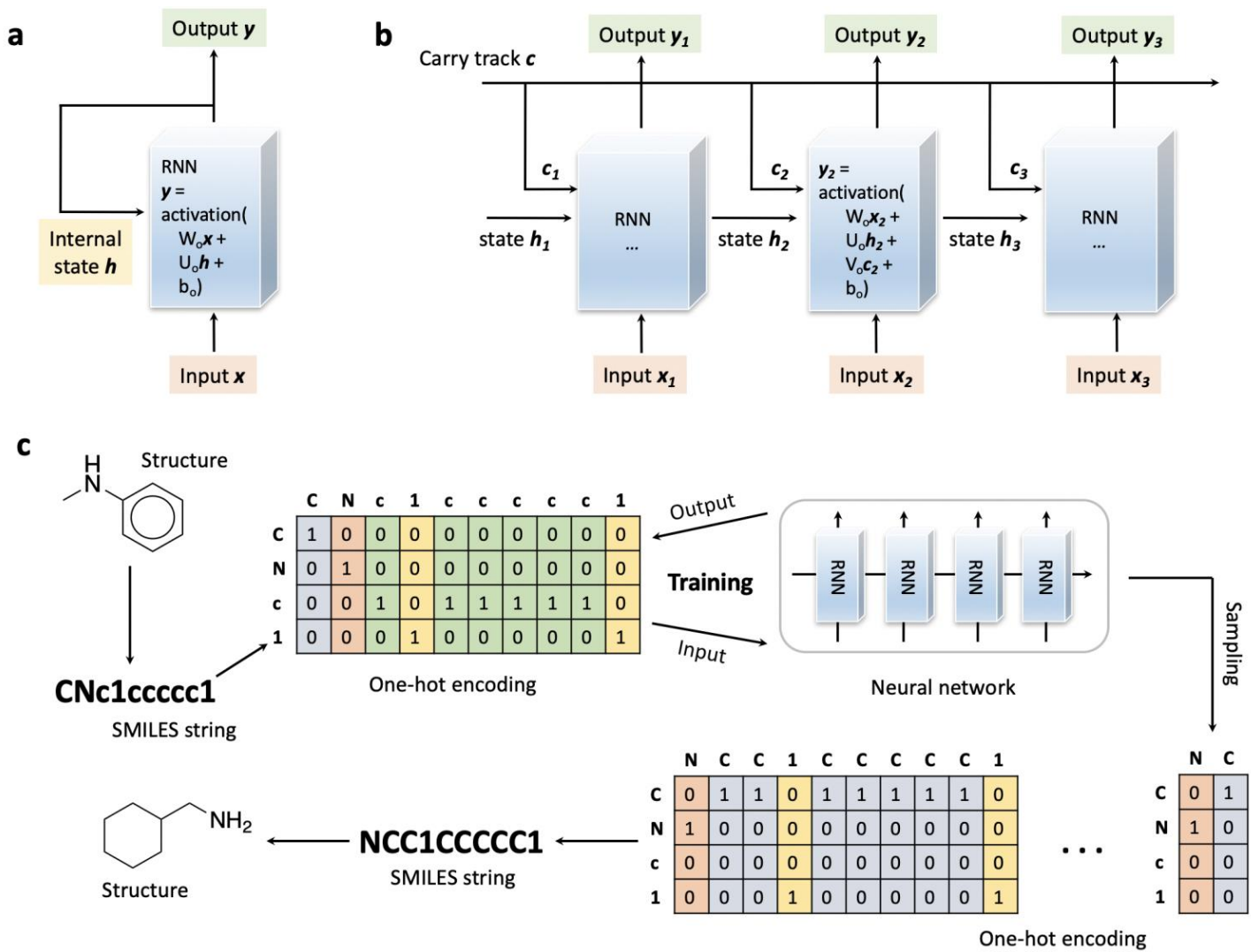

**Figure 2.** The RNN, the LSTM, and their application in generative chemistry. a. The schematic illustration of the RNN, the neural network with an internal loop. b. The schematic illustration of data processing with the LSTM. c. The typical framework on building generative models applying RNN for molecules generation.

Representative case studies are discussed in this paragraph. All the case applications covered in this review are summarized in **Table 4**. Anvita Gupta et al. trained an LSTM-based generative model with transfer learning to generate libraries of molecules with structural similarity to known actives for PPARγ and trypsin[91]. The model was first trained with 550,000 SMILES strings of active compounds from ChEMBL and further fine-tuned with SMILES strings for 4,367 PPARγ ligands and 1,490 trypsin inhibitors. Among the valid generated molecules, around 90% are unique from the known ligands and are different from each other. The proposed model was assessed for fragment-based drug discovery as well. In fragment-based drug discovery, fragment growing is a strategy for novel compounds generation with the identified fragment lead. Substitutions can be added to the identified fragment with the consideration of pharmacophore features and proper physical-chemical properties to enhance the receptor-ligand interactions[92]. Instead of using the starting character to initiate the generative process, the SMILES string of the molecular fragment can be read and extended by calculating the probability of distribution for the next character. Marwin H. S. Segler et al. also reported their application of LSTM-based generative models for structure generation with transfer learning[93]. There was a good correlation between the generated structures and the molecules used for training. Notably, the complete *de novo* drug design cycle can be achieved with target prediction models for scoring. As the target prediction model can be a molecular docking algorithm or even robot synthesis and bio-testing system, the drug design cycle does not require known active compounds to start. Chemical Language Model (CLM) proposed by Michael Moret et al. is another example of applying LSTM-based generative models to work with chemical SMILES strings with transfer learning processes[94]. This approach enables the early stage molecular design in a low data regime. When it comes to real-world validation, Daniel Merk et al. published their prospective study with experimental evaluations[95]. Using the SMILES strings as the input, the LSTM-based generative model was trained and fine-tuned with the transfer learning process for the peroxisome proliferator-activated receptor. Five top-ranked compounds designed by the model were synthesized and tested. Four of them have nanomolar to low micromolar activities in cell-based assays. Besides using the LSTM algorithm, some other RNN architectures such as implementing Gated Recurrent Unit[96] (GRU) can also have promising applications. GRU layers work with the same principle as LSTM layers but may have less representational power. Shuangjia Zheng et al. developed a quasi-biogenic molecular generator with GRU layers[97]. As biogenic compounds and pharmaceutical agents are biologically relevant, over 50% of existing drugs result from drug discovery campaigns starting with biogenic molecules. Their generative model is an effort to explore greater biogenic diversity space. Similarly, focused compound libraries can be constructed with transfer learning processes.

## 5. GENERATIVE CHEMISTRY WITH THE VARIATIONAL AUTOENCODER (VAE)

The principle aim of an autoencoder (AE) is to construct a low-dimensional latent space of compressed representations that each element can be reconstructed to the original input (**Fig. 3a**). The module that maps the original input data, which is in high-dimension, to a low-dimensional representation is called the encoder, while the module that realizes the mapping and reconstructs the original input from the low-dimensional representation is called the decoder[41, 98]. The encoder and the decoder are usually neural networks with RNN and CNN architectures as SMILES strings and molecular graphs are commonly used molecular representations. With the molecular representations calculated, a typical data processing procedure with AE on molecule generation starts with encoding the input into a low-dimensional latent space. Within the latent space, the axis of variations from the input can be encoded. Using the variation of molecular weight (M.W.) as an example, while in practice the features learned can be highly abstractive as the M.W. is used here for simplified illustration, the points along this axis are embedded representations of compounds with different M.W. These variations are termed concept vectors. With an identified vector, it makes the molecular editing possible by exploring the representations in a relevant direction. The encoded latent space with compressed representations can then be sampled with the decoder to map them back to molecular representations. Novel structures alongside the original input can be expected.

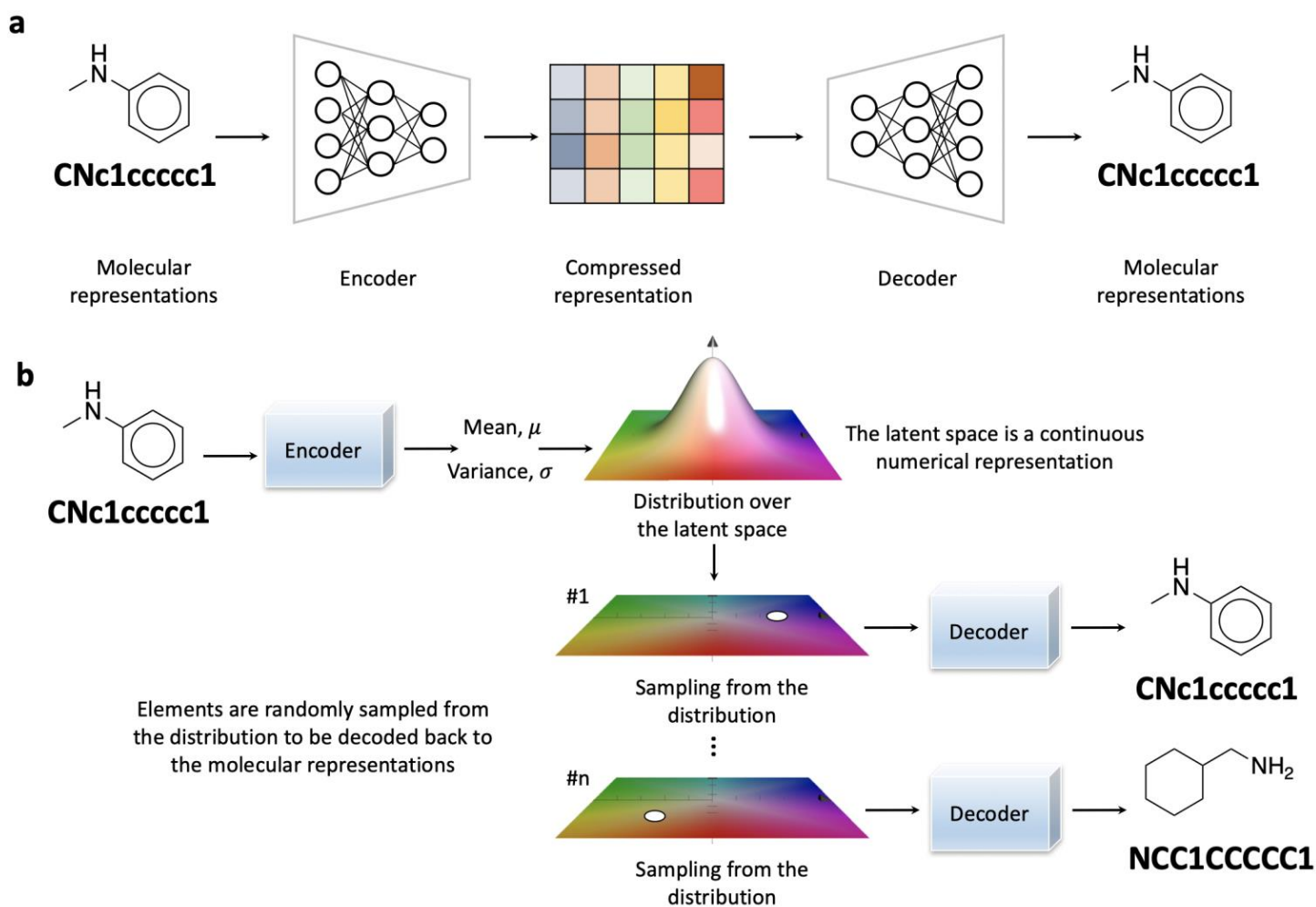

**Figure 3. The autoencoder and the variational autoencoder. a. An autoencoder encodes input molecules into compressed representations and decodes them back. b. A variational autoencoder maps the molecules into the parameters of a statistical distribution as the latent space is a continuous numerical representation.**

The concept of VAE was first proposed by Kingma and Welling at the end of 2013[99, 100]. This technique quickly gained popularity in building robust generative models for images, sounds, and texts[101-103]. The AE compresses a molecule $x$ into a fixed code in the continuous latent space $z$, and trends to summarize the explicit mapping rules as the number of adjustable parameters is often much more than the number of training molecules. These explicit rules make the decoding of random points in the continuous latent space challenging and sometimes impossible[26]. Instead, VAE maps the molecules into the parameters of a statistical distribution (**Fig. 3b**). With $p(z)$ describing the distribution of prior continuous latent space, the probabilistic encoding distribution is $q_\varphi(z|x)$ and the probabilistic decoding distribution is $p_\theta(x|z)$. The training iterations with back propagation will gradually optimize the parameters of both $q_\varphi(z|x)$ and $p_\theta(x|z)$. VAE is fundamentally a latent variable model $p(x,z) = p_\theta(x|z)p(z)$. The stochasticity of the training process enables the latent space to encode valid representations, which further results in a structured latent space[100]. Both the reconstruction loss and the regularization loss are often used for parameter optimization during the training process. The reconstruction loss evaluates whether the decoded samples match the input while the regularization loss investigates whether the latent space is overfitting to the training data.

Applications of VAE for generating chemical structures started in 2016 as Rafael Gómez-Bombarelli et al. developed a VAE-based automatic chemical design system[104]. In their practice, the ZINC database and QM9 dataset were referred to as the sources for collecting molecules. The QM9 dataset archives small molecules following three rules: (1) no more than 9 heavy atoms, (2) with 4 distinct atomic numbers, and (3) with 4 bond types. Canonical SMILES strings were calculated as the molecular representation. The encoder maps input SMILES strings into the continuous real-valued vectors, and the decoder reconstructs molecular representations from these vectors. The encoder was formed with three convolutional layers and one fully connected dense layer while the decoder contained three GRU layers. The architectures of CNNs and RNNs were compared for string encoding and convolutional layers achieved superior performance. The last layer of the decoder would report a probability distribution for characters of the SMILES string at each position. This stochastic operation allowed the same point in the latent space to have different decoded outcomes. Besides, they added one additional module for property prediction. An MLP was jointed to predict the property values from the continuous representation created by the encoder in order to optimize the desired properties for the new molecules. Thomas Blaschke et al. tested various generative AE models including VAE for compound design targeting dopamine receptor 2 (DRD2)[105]. Their study showed that the generated latent space preserved the chemical similarity principles. The generated molecules similar to known active compounds can be observed. In their VAE model, CNN layers were used for the encoder for pattern recognition and the RNN layers of GRU cells were adapted for the decoder. The ChEMBL database functioned as the data source for molecular structures. Canonical SMILES were prepared as the molecular representation. Similarly, an SVM classification model trained with extensive circular fingerprint (ECFP) of active and inactive DRD2 ligands was integrated to investigate the newly generated molecules. Boris Sattarov et al. combined a sequence-to-sequence VAE model with generative topographic mapping (GTM) for molecular design[106]. Both the encoder and the decoder were RNN models containing two LSTM layers in their practice. SMILES strings with one-hot encoding for molecules from the ChEMBL database were prepared prior to the training. Their GTM module contributed to the selection of sampling points in the VAE latent space, which facilitated the generation of a focused library of compounds with desired properties.

Besides the use of SMILES strings, molecular graphs have also been applied as a type of molecular representation to feed the VAE models. Bidisha Samanta et al. proposed NeVAE, a VAE-based compound generative model employing molecular graphs[107]. The molecular structures are usually not grid-like and come with an inconsistent number of nodes and edges, which impedes the use of molecular graphs as representations. In their work, the molecular graphs were prepared for drug-like compounds collected from the ZINC database and QM9 dataset. The nodes and edges in the graph represent atoms and bonds respectively. The node features are types of atoms with one-hot encoding and the edge weights are bond types (saturated bonds, unsaturated double/triple bonds, etc.). The purpose of training is to enable the VAE to create credible molecular graphs including node features and edge weights. Another example is GraphVAE. Martin Simonovsky et al. proposed GraphVAE to facilitate the compound design using molecular graphs[108]. Their central hypothesis was to decode a probabilistic fully-connected graph in which the existence of nodes, edges, and their attributes are independent random variables. The encoder was a feed-forward network with convolutional layers and the architecture for the decoder was an MLP. The model training and evaluation involved the molecules from the ZINC database and QM9 dataset. Some other generative applications can switch the topic to lead optimizations with methods such as scaffold hopping, substitutions design, and fragment-based approaches. One example is the DeLinker which was proposed by Fergus Imrie et al. to incorporate two fragments into a new molecule[109]. This method is VAE-based, using molecular graphs as the input. The design process heavily relied on 3D structural information that considers relative distance and orientation between the starting fragments.

## 6. GENERATIVE CHEMISTRY WITH THE ADVERSARIAL AUTOENCODER (AAE)

The architecture of the AAE is comparatively similar to the VAE except the appending of the additional discriminator network[110]. An AAE trains three modules, an encoder, a decoder, and a discriminator (**Fig. 4**). The encoder learns the input data and maps the molecule into the latent space following the distribution of $q_\varphi(z|x)$. The decoder reconstructs molecules through sampling from the latent space following the probabilistic decoding distribution of $p_\theta(x|z)$. And the discriminator distinguishes the distribution of the latent space $z \sim q_\varphi(z)$ from the prior distribution $z' \sim p(z)$. During the training iterations, the encoder is modified consistently to have the output, $q_\varphi(z|x)$, follow a specific distribution, $p(z)$, in an effort to minimize the adversarial cost of the discriminator. A simplistic prior, like Gaussian distribution, is assumed in VAE, while alternative priors can exist in real-world practices[111]. The AAE architecture with the additional discriminator module demonstrates improved adaptability.

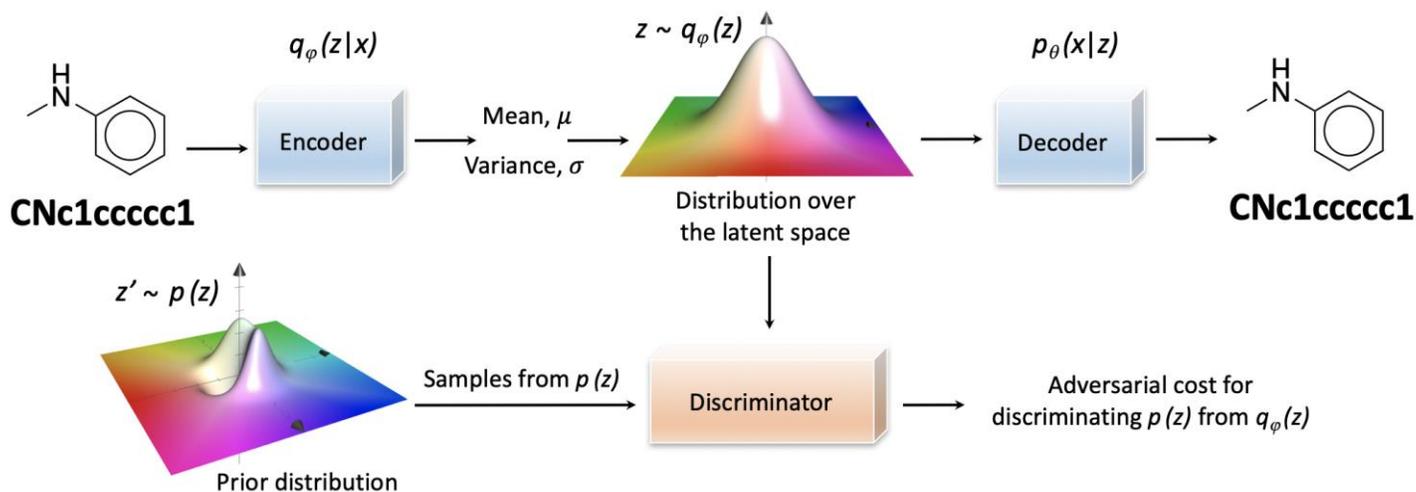

**Figure 4. The illustrated architecture of an adversarial autoencoder. A discriminator network is appended to calculate the adversarial cost for discriminating *p(z)* from *q$_\varphi$(z)*. As a result, the outcome latent space from the encoder is driven to follow the prior distribution.**

Thomas Blaschke et al. summarized a three-step training process in their compound design practice with AAE: (1) The simultaneous training of both the encoder and the decoder to curtail the reconstruction loss of the decoder; (2) The training of the discriminator to distinguish the distribution of the latent space, $q_\varphi(z)$, from the prior distribution *p(z)* effectively; (3) The training of the encoder to minimize the adversarial cost for discriminating *p(z)* from $q_\varphi(z)$[105]. The training iterations continue until the reconstruction loss converges. Artur Kadurin et al. proposed the method of using a generative adversarial autoencoder model to identify fingerprints of new molecules with potential anticancer properties[111]. The input molecules come from a small data set of compounds profiled on the MCF-7 cell line. The MACCS fingerprints were used as the molecular representation and two fully connected dense layers with different dimensions were used as the network architecture for the encoder, decoder, and the discriminator. One notable modification in this study was the removal of the batch normalization layers for the discriminator. Batch normalization is an optimization method that reduces the covariance shift among the hidden units and allows each layer to learn more independently. In the authors' opinion, the noise from the generator can be masked into target random noise with the batch normalization layers, which prohibits the training of the discriminator. As each bit of the MACCS fingerprints represents certain substructure features, the learned structural information by machine can be beneficial to the design of chemical derivatives for identified leads. Daniil Polykovskiy et al. reported their work on building a conditional AAE for molecule design targeting Janus kinase 3 (JAK3)[112]. The contributions from a set of physical-chemical properties including bioactivity, solubility, and synthesizability were considered and the model was conditioned to produce molecules with specified properties. Clean lead molecules were collected from the ZINC database and encoded as SMILES strings. The LSTM layers are adapted for building the encoder and the decoder networks. Both *in silico* method (molecular docking) and *in vitro* assay (inhibition of JAK2 and JAK3) were conducted as the evaluation for the newly generated molecules. Rim Shayakhmetov et al. reported a bidirectional AAE model that generates molecules with the capacity of inducing a desired change in gene expression[113]. The model was validated using LINCS L1000, a database that collects gene expression profiles. The molecular structures *x* and induced gene expression changes *y* contributed to a joint model *p(x,y)*. In this specific conditional task, there is no direct association between *x* and *y* as certain changes at the gene expression are irrelevant to the drug-target interactions. The proposed bidirectional AAE model then learned the joint distribution and decomposed objects into shared features, exclusive features to *x*, and exclusive features to *y*. Therefore, the discriminator that divides the latent representations into shared and exclusive sections was constructed to secure the conditional generation to be consequential.

**Table 4. Representative applications of generative chemistry covered in this review**

| # | Generative architecture | Neural networks involved | Data source | Molecular representation | Note | Ref. |
|---|---|---|---|---|---|---|
| 1 | RNN | LSTM | ChEMBL | SMILES | The application was extended to fragment-based drug design. | 91 |
| 2 | RNN | LSTM | ChEMBL | SMILES | The design-synthesis-test cycle was simulated with target prediction models for scoring. | 93 |
| 3 | RNN | LSTM | ChEMBL | SMILES | A chemical language model (CLM) in low data regimes. | 94 |
| 4 | RNN | LSTM | ChEMBL | SMILES | A prospective application with experimental validations of top-ranking compounds. | 95 |
| 5 | RNN | GRU | ZINC ChEMBL | SMILES | The generative model explored greater biogenic diversity space. | 97 |
| 6 | VAE | Encoder: CNN Decoder: GRU | ZINC QM9 | SMILES | An MLP model was jointed to predict property values. | 104 |
| 7 | VAE | Encoder: CNN Decoder: GRU | ChEMBL | SMILES | An SVM classification model was added to evaluate the outcome. | 105 |
| 8 | VAE | Encoder: LSTM Decoder: LSTM | ChEMBL | SMILES | A sequence-to-sequence VAE model was combined with generative topographic mapping (GTM) for molecular design. | 106 |
| 9 | VAE | Encoder: CNN Decoder: CNN | ZINC QM9 | Molecular graph | The nodes and edges in the graph of NeVAE represent atoms and bonds respectively. | 107 |
| 10 | VAE | Encoder: CNN Decoder: MLP | ZINC QM9 | Molecular graph | The central hypothesis of GraphVAE was to decode a probabilistic fully-connected graph. | 108 |

| 11 | VAE | Encoder: GGNN[#]<br>Decoder: GGNN | ZINC<br>CASF[*] | Molecular graph | DeLinker was designed to incorporate two fragments into a new molecule. | 109 |
| 12 | AAE | Encoder: MLP<br>Decoder: MLP<br>Discriminator: MLP | MCF-7[^] | MACCS fingerprints | Fingerprints cannot be directly converted to structures but can provide certain substructure information. | 111 |
| 13 | AAE | Encoder: LSTM<br>Decoder: LSTM<br>Discriminator: MLP | ZINC | SMILES | The generated molecules targeting JAK3 were evaluated with *in silico* and *in vitro* methods. | 112 |
| 14 | AAE | Encoder: GRU<br>Decoder: GRU<br>Discriminator: MLP | LINCS[&]<br>ChEMBL | SMILES | The combination of molecules and gene expression data were analyzed. | 113 |
| 15 | GAN | Discriminator: CNN<br>Generator: LSTM | ZINC | SMILES | Sequence generation with objective-reinforced generative adversarial networks (ORGAN). | 114 |
| 16 | GAN | Discriminator: MLP<br>Generator: MLP | ZINC | Molecular graph | The model operated in the latent space trained by the Junction Tree VAE. | 115 |
| 17 | GAN | Discriminator: MLP<br>Generator: MLP | LINCS[&] | SMILES | The compound design was connected to the systems biology. | 116 |
| 18 | GAN | Encoder: LSTM<br>Decoder: LSTM<br>Discriminator: MLP<br>Generator: MLP | ChEMBL | SMILES | The concept of the autoencoder and the generative adversarial network was combined to propose a latentGAN. | 117 |

[#]GGNN represents the gated graph neural network. [*]CASF is also known as PDBbind core set. [^]MCF-7 represents a small data set of compounds profiled on the MCF-7 cell line. [&]LINCS represents the LINCS L1000 dataset that collects gene expression profiles.

## 7. GENERATIVE CHEMISTRY WITH THE GENERATIVE ADVERSARIAL NETWORK (GAN)

The architecture of the convolutional neural network[43] (CNN) is briefly covered in this section as the convolutional layers are widely used in GAN modeling. The implementation of convolutional layers can also be found in case studies discussed above among autoencoder models. A convolutional layer does not learn an input globally but focuses on the local pattern within a receptive field, the kernel (**Fig. 5a**). The low-level patterns learned in a prior layer can then be concentrated on the high-level features at the subsequent layers[118, 119]. This characteristic allows the CNN to learn and summarize abstract patterns with complexity. Another characteristic that comes out from the local pattern learning is that the learned features can be recognized anywhere[118]. It enables the CNN to process input data with efficiency and powerfulness even with a smaller number of input sample representations. Meanwhile, multiple feature maps (filters) can be stacked to encode different aspects of the input data. Applying several filters capacitates a CNN model to detect distinct features anywhere among the input data. The pooling operation on the other hand subsamples the feature map to reduce the number of parameters and eventually, the computational load[120]. Using a max-pooling layer as one example, only the max input value in that pooling kernel will be kept. Alongside with dropout layers and regularization penalties, the pooling layers also contribute to confronting the overfitting issues. Putting together, the convolutional layers, pooling layers, and dense layers are carefully selected and arranged to construct a sophisticated CNN architecture.

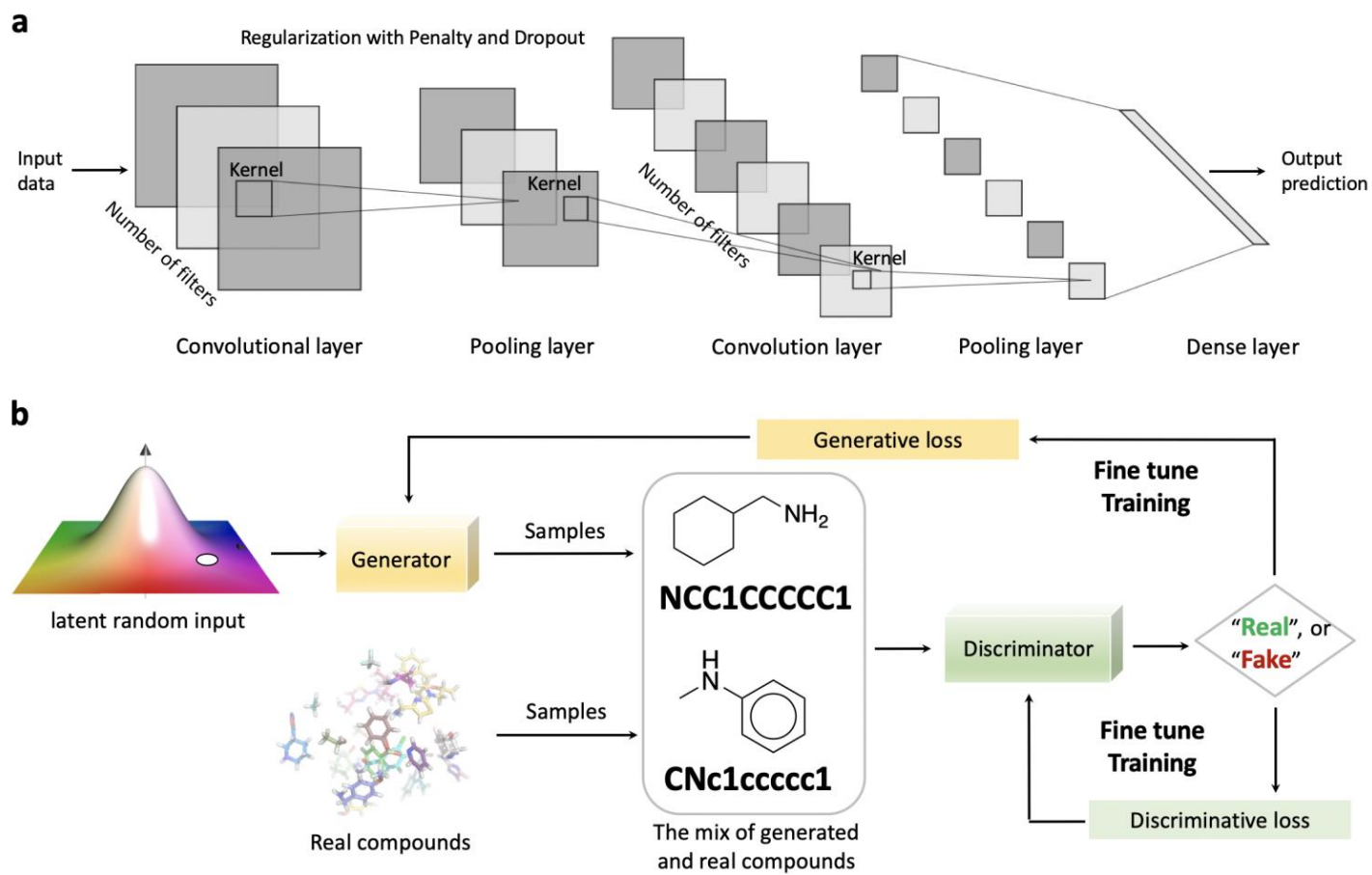

**Figure 5. Sample architecture of the convolutional neural network and the framework of a generative adversarial network. a. The careful selection and arrangement of convolutional layers, pooling layers, and dense layers, etc. constitute a convolutional neural network. b. The generative adversarial network comprises two modules, the generator and the discriminator. Both the generative loss and discriminative loss are monitored during the training process.**

The concept of the GAN was first raised by Ian Goodfellow in 2014[46]. The method quickly gained popularity on generative tasks regarding image, video, and audio processing and related areas[121-123]. Two models, the discriminator and the generator are trained iteratively and simultaneously during the adversarial training process[63]. The discriminator is supposed to discover the hidden patterns behind the input data and to make accurate discrimination of the authentic data from the ones generated by the generator. The generator is trained to keep proposing compelling data to fool the well-trained discriminator by consistently optimizing the data sampling process. The training process is a zero-sum noncooperative game with the purpose of achieving the Nash equilibrium by the discriminator and the generator. In generative chemistry, the generator generates SMILES strings, molecular graphs, or fingerprints, depending on the selection of the molecular representation, using the latent random inputs (**Fig. 5b**). The generated molecules are mixed with the samples of real compounds to feed the discriminator after correct labeling. The discriminative loss is calculated to evaluate whether the discriminator can distinguish the real compounds from the generated ones, while the generative loss is computed to assess whether the generator can fool the discriminator by generating undistinguishable molecules. The constringency of both loss functions after the iterative training indicates that even a well-established discriminator can be misled to classify generated molecules as real, which further reflects that the generator has learned and accumulated authentic data patterns to create captivating compounds. However, it is worth mentioning that the simultaneous optimization of both loss functions is challenging as the instability can lead to the gradient of one part instead of both being favored (results in a stronger discriminator or generator, but not both). Another limitation may come from the restricted chemical space that is being covered by the generated molecules[64]. To confront the discriminator and minimize the generative loss, the generator can only explore a limited chemical space defined by the real compounds.

Gabriel Guimaraes et al. presented a sequence-based GAN framework termed objective-reinforced generative adversarial network (ORGAN)[114] that includes domain-specific objectives to the training process besides the discriminator reward. The discriminator drove the generated samples to follow the distribution of the real data and the domain-specific objectives secured that the traits maximizing the specific heuristic would be selected. The drug-like and nondrug-like molecules were collected from ZINC databases. SMILES strings were calculated as the molecular representations. A CNN model was designed as the discriminator to classify texts, and an RNN model with LSTM units was used as the generator. Łukasz Maziarka et al. introduced Mol-CycleGAN for derivatives design and compound optimization[115]. The model could generate structures with high similarity to the original input but improved values on considered properties. Molecular graphs of compounds extracted from the ZINC database were used as the molecular representation. The model operated in the latent space trained by the Junction Tree VAE. Dense layers and fully connected residual layers constituted the generator and the discriminator. Oscar Méndez-Lucio et al. reported a GAN model to connect the compound design with systems biology[116]. They have shown that active-like molecules can be generated given that the gene expression signature of the selected target is supplied. The architectures of both the discriminator and the generator were composed with dense layers. There were two stages of training: in stage I, the random noise was taken as the input, while in stage II, the output from stage I and the gene expression signature were taken. Oleksii Prykhodko et al. combined the concept of AE with GAN and proposed a latent vector-based GAN model[117]. A heteroencoder mapped one-hot encoded SMILES strings into the latent space and the generator and discriminator would directly use the latent vector to focus on the optimization of the sampling process. A pre-trained heteroencoder was then used to transfer the generated vectors back to molecular structures. Both general drug-like compounds and target-biased molecules were generated as applications of the method.

## 8. CONCLUSION AND FUTURE PERSPECTIVES

Besides the successful generative chemistry stories described above, challenges and opportunities can be found at the following four aspects: (1) the synthetic feasibility of the generated structures, (2) the alternative molecular representations that can better portray a structure, (3) the generation of macro-molecules, and (4) the close-loop automation in combination with experimental validations. Wenhao Gao et al. pointed out that generative models can propose unrealistic molecules even with high performance scores on quantitative benchmarks[124]. Some existing methods of evaluating the synthesizability are based on synthetic routes and molecular structural data, which require heuristic definition to be complex and comprehensive[125], while the change of one single functional group to a scaffold can cause a distinctive synthetic pathway. The ignorance of synthesizability turns out to be an eminent hindrance of connecting generative models with medicinal chemistry synthesis. The molecular representations such as SMILES strings and molecular fingerprints serve well on describing small molecules at the current stage. However, it will be appealing if the novel representations can be designed to also consider three-dimensional geometry data. Chiral compounds may exhibit divergent activities to the biological system[126], and even the conformational change of the same small molecule can alter the receptor-ligand interactions. The case studies that deployed molecular graphs as the representation illustrate the benefits of working with structures directly[107-109, 115]. The extended consideration of bond type, length, and angles improves the performance of feature extraction on spatial patterns. Peptides possess superior advantage among protein subtype selectivity. The strategy of developing antibodies and peptides as therapeutic agents draws increasing attention from both the academia and industry. Deep learning is data-driven research. Current generative chemistry applications mainly focus on the design of small molecules as there is increased availability of accessing chemical data[127]. As the construction of protein-related databases is rising, the attempts of *de novo* protein generation are expected[128]. Better representations are certainly required for describing protein, as the folding and its conformation are even more critical to determine the functionality. Lastly, it is noteworthy of how to integrate the generative chemistry into the drug design framework to close the loop of this automated process. Marwin H. S. Segler et al. mentioned a design-synthesis-test cycle in their application of using the RNN model to generate molecules[93]. Ideally, the HTS will first recognize some hit compounds for a given target. The identified hits will contribute to the iterative training of a deep learning generative model for novel compounds generation, and a machine learning-based target prediction model for virtual classification. The top molecules will be synthesized and tested with biological assays. The true new actives can then be appended to the identified hits, which closes the loop.

In a nutshell, this paper reviewed the latest advances of generative chemistry that utilizes deep learning generative models to expedite the drug discovery process. The review starts with a brief history of AI in drug discovery to outline this emerging paradigm. Commonly used chemical databases, molecular representations, and operating sources of cheminformatics and machine learning are covered as the infrastructure. The detailed discussions on RNN, VAE, AAE, and GAN are centered, which is followed by future perspectives. As a fast-growing area of research, we are optimistic to expect a boosting number of studies on generative chemistry. We are probably at the corner of an upcoming revolution of drug discovery in the AI era, and the good news is that we are witnessing the change.

## 9. AUTHOR INFORMATION

**Corresponding author**

Author to whom correspondence should be addressed: Xiang-Qun Xie

Notes

The authors declare no competing financial interest.

## 10. ACKNOWLEDGEMENTS

Authors would like to acknowledge the funding support to the Xie laboratory from the NIH NIDA (P30 DA035778A1) and DOD (W81XWH-16-1-0490).